\documentstyle[epsf]{aipproc}

\begin{document}

%remove for camera-ready copy
\begin{flushright}UMN-D-99-2 \\ August 1999 \end{flushright}

\title{Pauli--Villars regularization in DLCQ%
%remove footnote for camera-ready copy
\footnote{To appear in the proceedings of 
the Eleventh International Light-Cone Workshop
on New Directions in Quantum Chromodynamics,
Kyungju, Korea, June 20-25, 1999.}
}

\author{John R. Hiller}
\address{Department of Physics \\
University of Minnesota Duluth \\
Duluth Minnesota 55812}

\maketitle

\begin{abstract}
  Calculations in a (3+1)-dimensional model indicate 
that Pauli-Villars regularization can be combined with 
discrete light-cone quantization (DLCQ) to solve at 
least some field theories nonperturbatively.  Discrete 
momentum states of Pauli-Villars particles are included 
in the Fock basis to automatically generate needed 
counterterms; the resultant increase in basis size is 
found acceptable.  The Lanczos algorithm is used to 
extract the lowest massive eigenstate and eigenvalue of 
the light-cone Hamiltonian, with basis sizes ranging up 
to 10.5 million.  Each Fock-sector wave function is
computed in this way, and from these one can obtain
values for various quantities, such as average 
multiplicities and average momenta of constituents, 
structure functions, and a form-factor slope.
\end{abstract}

\section*{Introduction}

Field-theoretic calculations of bound-state properties,
such as those one would like to do for quantum
chromodynamics (QCD), require regularization of
infinities and renormalization of parameters.  As 
a way of providing a systematic regularization of
ultraviolet infinities we study \cite{PV1,PV2}
the use of Pauli--Villars (PV) regularization \cite{PauliVillars}
in the context of discrete light-cone quantization 
(DLCQ) \cite{PauliBrodsky,DLCQreview}.  Renormalization
is accomplished by adjusting bare parameters to fit
selected state properties with ``data.''  The problem
to be solved is then a bound-state eigenvalue problem,
which includes PV constituents, combined with
renormalization conditions.  The couplings of the
PV constituents are chosen to produce desired
cancellations in perturbation theory.

We have tested these ideas for two related 
(3+1)-dimensional Hamiltonians
\cite{PV1,PV2}.  The first \cite{PV1} was constructed
to have an analytic solution, in analogy with the
equal-time model of Greenberg and Schweber \cite{GreenbergSchweber}.
The second Hamiltonian \cite{PV2} is a generalization
of the first which assigns proper light-cone energies
to all particles, but does not have an analytic solution.
Both Hamiltonians are distantly related to Yukawa theory,
in that a fermion field acts as a source and sink for
bosons.  The second Hamiltonian and some of the results
obtained will be described here.  Work on direct application
to Yukawa theory is in progress.

The choice of light-cone coordinates ($x^\pm =t\pm z$,
$\bf{x}_\perp=(x,y)$) \cite{Dirac,DLCQreview} is 
driven by important advantages, which include 
kinematical boosts, a simple vacuum, and well-defined
Fock-state expansions with no disconnected pieces.
The latter two derive from the positivity of the
longitudinal light-cone momentum $p^+=E+p_z$.
When momenta are discretized \cite{PauliBrodsky} this 
positivity brings the additional advantage of a
finite limit on the number of constituents.

\section*{Model Eigenvalue Problem}

The bound-state eigenvalue problem is 
$H_{\rm LC}\Phi_\sigma=M^2\Phi_\sigma$,
where $H_{\rm LC}=P^+P^-$ is known as the light-cone Hamiltonian,
$P^+$ is the longitudinal momentum operator, and $P^-$ is the 
generator for evolution in light-cone time.  We work in the frame
with no net transverse momentum and in a basis diagonal in $P^+$.

The Hamiltonian that we consider is
\begin{eqnarray}
H_{\rm LC}&=&\int\frac{dp^+d^2p_\perp}{16\pi^3p^+}
     (\frac{M^2+p_\perp^2}{p^+/P^+}+M'_0p^+/P^+)
   \sum_\sigma b_{\underline{p}\sigma}^\dagger b_{\underline{p}\sigma} 
\nonumber \\
  & & +\int\frac{dq^+d^2q_\perp}{16\pi^3q^+}
       \left[\frac{\mu^2+q_\perp^2}{q^+/P^+}
            a_{\underline{q}}^\dagger a_{\underline{q}}
           + \frac{\mu_1^2+q_\perp^2}{q^+/P^+}
                a_{1\underline{q}}^\dagger a_{1\underline{q}} \right]
\nonumber  \\
   &  & +g\int\frac{dp_1^+d^2p_{\perp1}}{\sqrt{16\pi^3p_1^+}}
            \int\frac{dp_2^+d^2p_{\perp2}}{\sqrt{16\pi^3p_2^+}}
              \int\frac{dq^+d^2q_\perp}{16\pi^3q^+}
                \sum_\sigma b_{\underline{p}_1\sigma}^\dagger 
                         b_{\underline{p}_2\sigma}   \\
    &  & \rule{0.25in}{0mm}\times \left[
         a_{\underline{q}}^\dagger
         \delta(\underline{p}_1-\underline{p}_2+\underline{q})
        +a_{\underline{q}}
         \delta(\underline{p}_1-\underline{p}_2-\underline{q}) \right.
\nonumber \\
    &  & \rule{0.5in}{0mm} \left.
     +i a_{1\underline{q}}^\dagger
          \delta(\underline{p}_1-\underline{p}_2+\underline{q})
    +i a_{1\underline{q}}
       \delta(\underline{p}_1-\underline{p}_2-\underline{q}) \right]\,.
\nonumber 
\end{eqnarray}
The creation operators $b_{\underline{p}\sigma}^\dagger$,
$a_{\underline{q}}^\dagger$, and $a_{1\underline{q}}^\dagger$
are associated with fermion, boson, and PV boson fields,
respectively.  The corresponding masses are $M$, $\mu$,
and $\mu_1$.  Each operator depends on a light-cone
three-momentum such as $\underline{p}\equiv(p^+,p_x,p_y)$.  The
nonzero commutation relations are
\begin{eqnarray}
\left\{b_{\underline{p}\sigma},b_{\underline{p}'\sigma'}^\dagger\right\}
     &=&16\pi^3p^+\delta(\underline{p}-\underline{p}')
                                   \delta_{\sigma\sigma'}\,, \\
\left[a_{\underline{q}},a_{\underline{q}'}^\dagger\right]
          &=&16\pi^3q^+\delta(\underline{q}-\underline{q}')\,, \;\;
\left[a_{1\underline{q}},a_{1\underline{q}'}^\dagger\right]
          =16\pi^3q^+\delta(\underline{q}-\underline{q}')\,.
\nonumber 
\end{eqnarray}
The structure of the Hamiltonian provides for emission and
absorption of bosons by the fermion, but no change in fermion
number.  We explore only the one-fermion sector.  
The particular form of the interaction causes the 
fermion mass counterterm to have an unusual momentum dependence.
The coefficient of this counterterm is finite because of
cancellations arranged by assigning an imaginary coupling
to the PV boson.

The state vector $\Phi_\sigma$ describes a dressed fermion
with spin $\sigma$.  Its Fock-state expansion is given by
\begin{eqnarray}
\Phi_\sigma&=&\sqrt{16\pi^3P^+}\sum_{n,n_1}
                  \int\frac{dp^+d^2p_\perp}{\sqrt{16\pi^3p^+}}
 \prod_{i=1}^n\int\frac{dq_i^+d^2q_{\perp i}}{\sqrt{16\pi^3q_i^+}}
 \prod_{j=1}^{n_1}\int\frac{dr_j^+d^2r_{\perp j}}{\sqrt{16\pi^3r_j^+}} \\
   &  & \times \delta(\underline{P}-\underline{p}
                     -\sum_i^n\underline{q}_i-\sum_j^{n_1}\underline{r}_j)
       \phi^{(n,n_1)}(\underline{q}_i,\underline{r}_j;\underline{p})
   \frac{1}{\sqrt{n!n_1!}}b_{\underline{p}\sigma}^\dagger
          \prod_i^n a_{\underline{q}_i}^\dagger 
             \prod_j^{n_1} a_{1\underline{r}_j}^\dagger |0\rangle \,,
\nonumber
\end{eqnarray}
with normalization 
$\Phi_\sigma^{\prime\dagger}\cdot\Phi_\sigma
=16\pi^3P^+\delta(\underline{P}'-\underline{P})$, which implies
\begin{equation} \label{eq:Norm}
1=\sum_{n,n_1}\prod_i^n\int\,dq_i^+d^2q_{\perp i}
                     \prod_j^{n_1}\int\,dr_j^+d^2r_{\perp j}
    \left|\phi^{(n,n_1)}(\underline{q}_i,\underline{r}_j;
           \underline{P}-\sum_i\underline{q}_i-\sum_j\underline{r}_j)
                                             \right|^2\,.
\end{equation}
To satisfy the eigenvalue condition, the Fock-sector wave functions
$\phi^{(n,n_1)}$ must solve the following coupled system: 
\begin{eqnarray} \label{eq:CoupledEqns}
\lefteqn{\left[M^2-\frac{M^2+p_\perp^2}{p^+/P^+}-M'_0p^+/P^+
  -\sum_i\frac{\mu^2+q_{\perp i}^2}{q_i^+/P^+}
                  -\sum_j\frac{\mu_1^2+r_{\perp j}^2}{r_j/P^+}\right]
                    \phi^{(n,n_1)}(\underline{q}_i,
                                       \underline{r}_j,\underline{p}) } 
\nonumber \\
& & =g\left\{\sqrt{n+1}\int\frac{dq^+d^2q_\perp}{\sqrt{16\pi^3q^+}}
              \phi^{(n+1,n_1)}(\underline{q}_i,\underline{q},
                     \underline{r}_j,\underline{p}-\underline{q})\right.
\\
& &\rule{0.5in}{0mm} +\frac{1}{\sqrt{n}}\sum_i\frac{1}{\sqrt{16\pi^3q_i^+}}
       \phi^{(n-1,n_1)}(\underline{q}_1,\ldots,\underline{q}_{i-1},
                            \underline{q}_{i+1},\ldots,\underline{q}_n,
                            \underline{r}_j,\underline{p}+\underline{q}_i)
\nonumber \\
& &\rule{0.5in}{0mm}+i\sqrt{n_1+1}
                     \int\frac{dr^+d^2r_\perp}{\sqrt{16\pi^3r^+}}
              \phi^{(n,n_1+1)}(\underline{q}_i,\underline{r}_j,
                               \underline{r},\underline{p}-\underline{r})
\nonumber \\
& &\rule{0.5in}{0mm}+
       \left.\frac{i}{\sqrt{n_1}}\sum_j\frac{1}{\sqrt{16\pi^3r_j^+}}
 \phi^{(n,n_1-1)}(\underline{q}_i,\underline{r}_1,
                       \ldots,\underline{r}_{j-1},
                 \underline{r}_{j+1},\ldots,\underline{r}_{n_1},
                   \underline{p}+\underline{r}_j) \right\}\,.
\nonumber 
\end{eqnarray}

The bare parameters $M'_0$ and $g$ are determined by fitting
\mbox{$\langle :\!\!\phi^2(0)\!\!:\rangle
   \equiv\Phi_\sigma^\dagger\!:\!\!\phi^2(0)\!\!:\!\Phi_\sigma$}
and $M^2$ to chosen values.  The quantity 
$\langle :\!\!\phi^2(0)\!\!:\rangle$ was selected for ease of
computation; it can be computed from a form similar to the
normalization sum (\ref{eq:Norm}): 
\begin{eqnarray}
\langle :\!\!\phi^2(0)\!\!:\rangle
        &=&\sum_{n=1,n_1=0}\prod_i^n\int\,dq_i^+d^2q_{\perp i} 
                \prod_j^{n_1}\int\,dr_j^+d^2r_{\perp j}
   \\
    & & \rule{0.25in}{0mm} 
     \times \left(\sum_{k=1}^n \frac{2}{q_k^+/P^+}\right)
           \left|\phi^{(n,n_1)}(\underline{q}_i,\underline{r}_j;
       \underline{P}-\sum_i\underline{q}_i-\sum_j\underline{r}_j)
          \right|^2\,.    \nonumber
\end{eqnarray}

The renormalization conditions that determine $M'_0$ and $g$
are then solved simultaneously with the eigenvalue problem
(\ref{eq:CoupledEqns}).  In practice this is done by 
rearranging (\ref{eq:CoupledEqns}) into an eigenvalue problem
for $1/g$ and simultaneously solving for $M'_0$ in a single nonlinear
equation where $\langle :\!\!\phi^2(0)\!\!:\rangle$ is equal to
a fixed value.  The simultaneous solution is done by iterative 
means.

Once the wave functions $\phi^{(n,n_1)}$ have been
obtained, they can be used to compute various 
quantities, such as the boson structure function
\begin{eqnarray}
f_B(y)&\equiv&\sum_{n,n_1}\prod_i^n\int\,dq_i^+d^2q_{\perp i}
                     \prod_j^{n_1}\int\,dr_j^+d^2r_{\perp j} 
     \sum_{i=1}^n\delta(y-q_i^+/P^+)  \\
&&\rule{0.5in}{0mm}\times
    \left|\phi^{(n,n_1)}(\underline{q}_i,\underline{r}_j;
        \underline{P}-\sum_i\underline{q}_i-\sum_j\underline{r}_j)
                   \right|^2          \nonumber
\end{eqnarray}
and the average boson multiplicity and momentum
\begin{equation}
\langle n_B\rangle=\int_0^1f_B(y)dy\,,\;\;
\langle y\rangle=\int_0^1 yf_B(y)dy\,.
\end{equation}
We also compute the slope of the fermion ``charge'' form factor
$F'(0)$, from an expression derived in Ref.~\cite{PV1}.

\section*{Numerical methods and results}

The coupled equations (\ref{eq:CoupledEqns}) are converted to
a finite matrix eigenvalue problem by applying the DLCQ
procedure \cite{PauliBrodsky}.  Integrals are approximated
by sums over discrete momentum values 
$(n\pi/L,n_x\pi/L_\perp,n_y\pi/L_\perp)$, and the transverse
range is limited by a cutoff $\Lambda^2$ such that
$(m_i^2+p_{\perp i})\leq\Lambda^2p_i^+/P^+$ for each
particle, with $m_i$ being its mass.  The length scales
$L$ and $L_\perp$ are associated with a light-cone 
coordinate box.  Bosons are assigned periodic boundary conditions
and the fermion is assigned an antiperiodic boundary condition
in the longitudinal direction.  The momentum integer $n$ is 
then even for bosons and odd for the fermion.

The total longitudinal momentum $P^+$ of the
dressed fermion defines an odd integer $K=P^+L/\pi$,
called the harmonic resolution \cite{PauliBrodsky}.
A longitudinal momentum fraction $x=p^+/P^+$ then 
reduces to a rational number $n/K$.  The positivity
of longitudinal momentum implies that $0<n\leq K$
and that the maximum number of constituents is of
order $K/2$.
The integers $n_x$ and $n_y$ range between $-N_\perp$
and $N_\perp$, with $N_\perp$ set to reach the limit 
imposed by the transverse cutoff for the one-boson
physical state.

Thus the discretization is determined by three
parameters: $K$, $N_\perp$, and $\Lambda^2$.  The
transverse scale $L_\perp$ is computed from these
as $L_\perp=\pi N_\perp\sqrt{2/(\Lambda^2-M^2-\mu^2)}$.
The longitudinal scale does not appear \cite{PauliBrodsky},
but the limit $L\rightarrow\infty$ is equivalent to
$K\rightarrow\infty$.  We therefore study the
limit where $K$, $N_\perp$, and $\Lambda^2$
all become large.  This recovers the continuum 
form of the theory, which is regulated by the
PV mass $\mu_1$, not by $\Lambda$.  We must then
also study the large $\mu_1$ limit.

Typical discretizations, such as $K=15$, $N_\perp=6$,
and $\Lambda^2=50\mu^2$, with $\mu_1^2=10\mu^2$
produce matrices with ranks on the order of
5 million.  The largest calculations carried out were
of rank 10.5 million, which required approximately
2 hours of cpu time on a single 4-processor node of
an IBM SP.  The diagonalization method used is the
Lanczos algorithm for complex symmetric matrices
\cite{Lanczos}.

Although the automatic truncation of particle number 
imposed by DLCQ can be sufficient, further truncation
can be made when the coupling is weak.  Such
truncation, typically to 4 bosons, 
was used to permit increased resolution
within fixed memory limits.  The validity of such
an approximation was checked by computing the 
contribution of individual Fock sectors to the
total norm.

\begin{figure}[b!]
%\vspace{3in} \special{eps:fb-all.eps y=3in} \vspace{0.3in} 
\centerline{\epsfxsize=\columnwidth \epsfbox{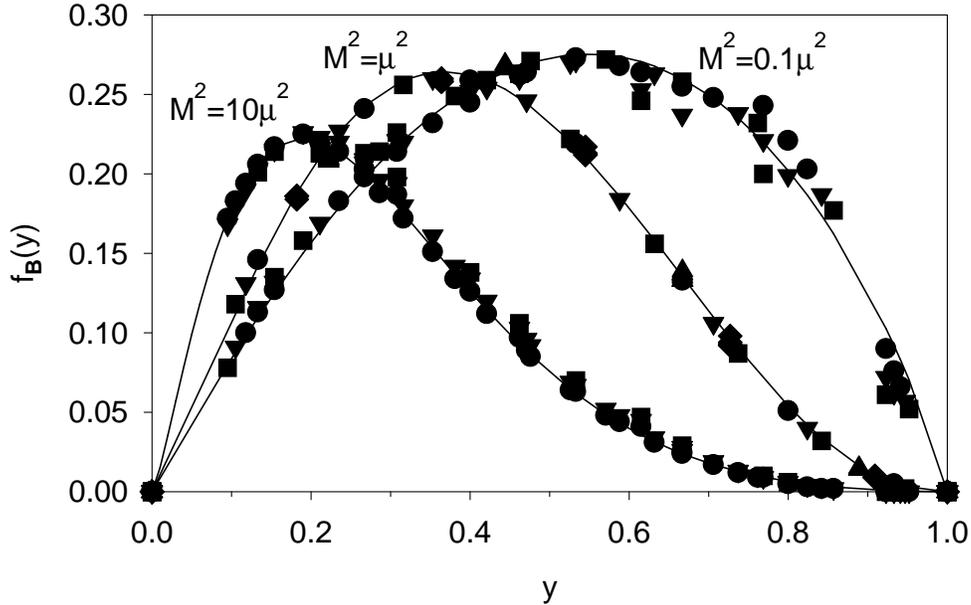} }
\caption{\label{fig:fB} The boson structure function $f_B$ at 
various numerical resolutions and mass values, 
with $\langle:\!\!\phi^2(0)\!\!:\rangle=1$,
$\Lambda^2=50\mu^2$, and $\mu_1^2=10\mu^2$. The solid lines are
parameterized fits of the form $Ay^a(1-y)^be^{-cy}$.} 
\end{figure}

Most quantities are remarkably insensitive to numerical
resolution.  This can be seen in Fig.~\ref{fig:fB} where
we display the boson structure function $f_B$ for 
various mass ratios $M/\mu$.  Different values of $K$
and $N_\perp$ generally do not yield significantly
different results.  Notice that a smaller mass ratio
is associated with a state where the boson 
constituents carry more of the momentum.

Values for a set of bound-state observables are given in
Table~\ref{tab:extrapolated}.
The quantity $|\phi_0|^2$ represents the probability
for the bare fermion state.
Each entry has been extrapolated from the numerical results
by fits to the form $\alpha+\beta/K^2+\gamma/N_\perp^2$.
To obtain the behavior of this form the use of weighting
factors \cite{PV1} in DLCQ sums is important.  The numerical
resolutions used ranged from 9 to 19 for $K$ and 5 to
as much as 10 for $N_\perp$; the larger values of $N_\perp$
were available only for smaller $K$.  Most observables
converge quickly with respect to the PV regulator mass
$\mu_1$.  Only $M'_0$ is
strongly dependent on $\mu_1$.  The form factor slope
is sensitive to the transverse resolution and range.

\begin{table}[t!]
\caption{\label{tab:extrapolated}
Extrapolated bare parameters and observables,
with $\langle:\!\!\phi^2(0)\!\!:\rangle=1$.
}
\begin{tabular}{ccccccccccc}
$(M/\mu)^2$       & 1 & 1 & 1 & 1  & 1  & 1  & 1  & 0.1 & 5  & 10  \\
$(\mu_1/\mu)^2$   & 5 & 5 & 5 & 10 & 10 & 20 & 20 & 10  & 10 & 10   \\
$(\Lambda/\mu)^2$ & 12.5 & 25 & 50 & 25 & 50 & 50 & 100 
                      & 50 & 100 & 100  \\
\hline
$g/\mu$           & 21.4 & 17.7 & 16.3 & 17.8 & 16.0 & 16.0 & 15.5 
                               & 15.1 & 18.1 & 19.0  \\
$M'_0/\mu^2$      & 1.26 & 1.10 & 1.10 & 1.48 &  1.4 & 1.8  & 1.9 
                               & 1.39 & 1.66 & 1.60  \\
\hline
$|\phi_0|^2$      & 0.82 & 0.83 & 0.84 & 0.85 & 0.86 & 0.87 & 0.87 
                                  & 0.83 & 0.89 & 0.90  \\
$-100\mu^2 F'(0)$ & 1.04 & 0.78 & 0.66 & 0.72 & 0.59 & 0.59 & 0.51 
                                  & 2.0 & 0.14 & 0.07  \\
$\langle n_B\rangle$ & 0.18 & 0.15 & 0.14 & 0.15 & 0.14 & 0.13 & 0.13 
                             & 0.16 & 0.10 & 0.09 \\
$\langle y\rangle$& 0.077 & 0.062 & 0.057 & 0.062 & 0.056 & 0.056 & 0.053 
                              & 0.073 & 0.032 & 0.024 \\
\end{tabular}
\end{table}

\section*{Summary}

This work shows that PV regularization is feasible
for DLCQ calculations.  The matrix size does increase
but not beyond the capacity of present-day machines
for the models considered.  More complicated theories
will require multiple computing nodes and message-passing
technology.  Use of multiple nodes is facilitated by
a natural block structure that arises in the matrix
due to limited coupling between Fock sectors.

Work on Yukawa theory in a single-fermion truncation
is now in progress.  The complications include additional
PV boson flavors and nontrivial spin dependence.
Quantum electrodynamics is perhaps the next logical
step.  QCD could also be considered in a broken
supersymmetric form that contains heavy particles
analogous to the Abelian PV particles introduced here.

\section*{Acknowledgments}
This work was done in collaboration with S.J. Brodsky and G. McCartor
and was supported in part by the Minnesota Supercomputing Institute
through grants of computing time and by the Department of Energy
contract DE-FG02-98ER41087.

\end{document}